\documentstyle[preprint]{ptptex}
\newcommand{\bra}[1]{\langle {#1} |}     %%
\newcommand{\ket}[1]{| {#1} \rangle}     %%
\newcommand{\bbra}[1]{\langle\!\langle {#1} |}     %%
\newcommand{\kket}[1]{| {#1} \rangle\!\rangle}     %%
\newcommand{\rbra}[1]{( {#1} |}     %%
\newcommand{\rket}[1]{| {#1} )}     %%
\newcommand{\maru}[1]{\breve{#1}} %%
\newcommand{\wtilde}[1]{\widetilde{#1}} %%
\title{%        %You can use \\ for explicit line-break
On the Multiboson Coherent State\\
in Deformed Boson Scheme
}
\author{%       %Use \sc for the family name
Atsushi {\sc Kuriyama}, 
Constan\c{c}a {\sc Provid\^encia}$^{*}$, \\
Jo\~ao da {\sc Provid\^encia}$^{*}$, Yasuhiko {\sc Tsue}$^{**}$ 
and Masatoshi {\sc Yamamura}
}

\inst{%         %Affiliation, neglected when [addenda] or [errata]
Faculty of Engineering, Kansai University, Suita 564-8680, Japan\\
$^{*}$Departamento de Fisica, Universidade de Coimbra, P-3000 Coimbra, 
Portugal\\
$^{**}$Physics Division, Faculty of Science, Kochi University, Kochi 780-8520, 
Japan
}

\recdate{%      %Editorial Office will fill in this.
}

\abst{%       %this abstract is neglected when [addenda] or [errata]
The treatment for the multiboson coherent state
is extended for the completeness of the formulation. 
The basic idea for the extension is presented in terms of expanding 
the boson space. The MYT boson mapping plays a central role. 
The resultant multiboson coherent state includes the state obtained 
in our previous 
treatment 
and the form suggests various multiboson coherent states 
are possible in 
the deformed boson scheme. 
}

\begin{document}

\maketitle

%\section{Introduction}

Recently, the present authors published four 
papers\cite{KPPTYI,KPPTYII,KPPTYIII} 
concerning the deformed boson scheme including conventional $q$-deformation 
in time-dependent variational method. 
The series of these papers consists of (I),\cite{KPPTYI} 
(II),\cite{KPPTYII} (III)\cite{KPPTYII} and the short note\cite{KPPTYIII} 
and, in this series, we discussed various aspects of the $q$-deformation 
appearing in the boson coherent state and its generalized form. 
For example, in (I),\cite{KPPTYI}
a possible form of the multiboson coherent state and its generalization 
were described. Its treatment is, in some sense, in close connection 
with that given by Penson and Solomon.\cite{Penson} 
However, the treatment in (I) is limited to a certain special boson subspace. 
In this sense, it is unsatisfactory and to make it complete is the aim 
of this paper. Further, the present paper gives 
a generalization of the treatment 
proposed by the present authors, which is related to a new boson 
realization in the Lipkin model.\cite{KPPYI,KPPYII}

In (I), a possible form of the multiboson coherent state, which we 
denote as $\kket{c_0}$, was defined in the boson space ${\wtilde B}$ 
composed by the boson operator $({\wtilde c}, {\wtilde c}^*)$ 
in the following form :
\begin{equation}\label{1}
\kket{c_0}=\left(\sqrt{\Gamma_0}\right)^{-1}
\sum_{n=0}^\infty \left(\sqrt{(mn)!}\right)^{-1}\gamma^n\kket{mn}\ .
\end{equation}
Here, $\gamma$ denotes a complex parameter. The quantity 
$\Gamma_0$ is the normalization constant given as 
\begin{equation}\label{2}
\Gamma_0=\sum_{n=0}^{\infty} \left[(mn)!\right]^{-1}(|\gamma|^2)^n \ .
\end{equation}
The state $\kket{mn}$ is expressed in terms of 
\begin{subequations}\label{3all}
\begin{eqnarray}
& &\kket{mn}=\left(\sqrt{(mn)!}\right)^{-1}({\wtilde c}^*)^{mn}\kket{0} \ , 
\qquad (n=0, 1, 2, \cdots) 
\label{3}\\
& &{\wtilde c}\kket{0}=0 \ . 
\label{3a}
\end{eqnarray}
\end{subequations}
The operator $({\wtilde c}^*)^m$ is the building block of the state 
$\kket{c_0}$ and, in this sense, we can call $\kket{c_0}$ the 
multiboson coherent state 
in which $m$ denotes an integer ($m=2, 3, \cdots$) characterizing 
the multiboson. 
Clearly, the set $\{\kket{mn}\}$ is 
an orthogonal but not a complete one : The set composes a subspace of 
${\wtilde B}$, which we call ${\wtilde B}_0$. 
%Needless to say, $m$ denotes 
%an integer $(m=2, 3, \cdots)$ which characterizes the multiboson. 

With the aim of investigating the structure of the deformation of 
the multiboson coherent state, in (I), we adopted the basic idea of the 
MYT boson mapping.\cite{MYT} 
For the image of the space ${\wtilde B}_0$, we prepare a boson 
space ${\hat B}_0$ constructed by the boson operator 
$({\hat c} , {\hat c}^*)$. The orthogonal set of ${\hat B}_0$ consists 
of the states 
\begin{subequations}\label{4all}
\begin{eqnarray}
& &\ket{n}=(\sqrt{n!})^{-1} ({\hat c}^*)^n \ket{0} , \qquad
(n=0, 1, 2, \cdots) 
\label{4}\\
& &{\hat c}\ket{0}=0 \ . 
\label{4a}
\end{eqnarray}
\end{subequations}
The set $\{\ket{n}\}$ is complete. Following the basic idea of the MYT 
boson mapping, we make the set $\{ \kket{mn} \}$ correspond to 
$\{ \ket{n} \}$ :
\begin{equation}\label{5}
\kket{mn} \sim \ket{n} \ , \qquad \hbox{\rm i.e.,}\qquad 
\ket{n}={\mib U}_0 \kket{mn} \ .
\end{equation}
The operator ${\mib U}_0$ plays a role of the mapping operator 
from ${\wtilde B}_0$, 
a subspace of ${\wtilde B}$, to ${\hat B}_0$ and it is given by 
\begin{equation}\label{6}
{\mib U}_0=\sum_{n=0}^{\infty} \ket{n}\bbra{mn} \ . 
\end{equation}
The operator ${\mib U}_0$ satisfies the relation 
\begin{equation}\label{7}
{\mib U}_0{\mib U}_0^\dagger = 1 \ , \qquad 
{\mib U}_0^\dagger {\mib U}_0 = {\wtilde P}_m\ . 
\qquad ({\wtilde P}_m = \sum_{n=0}^\infty \kket{mn}\bbra{mn} \ )
\end{equation}
Then, the image of $\kket{c_0}$ mapped from ${\wtilde B}_0$ to ${\hat B}_0$, 
which we denote $\ket{c_0}$, is given in the following form : 
\begin{eqnarray}
\ket{c_0}&=&{\mib U}_0 \kket{c_0} \nonumber\\
&=&\left(\sqrt{\Gamma_0}\right)^{-1} 
\sum_{n=0}^\infty f_m^0(n) (\sqrt{n!})^{-1}\gamma^n \ket{n} \ ,
\label{8}\\
f_m^0(n)&=&\sqrt{n!}\left(\sqrt{(mn)!}\right)^{-1} \ .
\label{9}
\end{eqnarray}
The form (\ref{8}) shows that $\ket{c_0}$ is deformed from the 
conventional boson coherent state by the function $f_m^0(n)$. 
This means that by changing $f_m^0(n)$ from the form (\ref{9}), we 
have various deformed states. The state $\ket{c_0}$ can be rewritten as 
\begin{eqnarray}
& &\ket{c_0}=\left(\sqrt{\Gamma_0}\right)^{-1}
\exp\left((\sqrt{m^m})^{-1}\gamma{\hat c}^* \left(\sqrt{F_m^0({\hat N}_c)}
\right)^{-1}\right) \ket{0} \ , 
\label{10}\\
& &F_m^0({\hat N}_c)=\prod_{p=1}^{m-1}
\left({\hat N}_c+p/m\right) = \left({\hat N}_c+1\right)^{-1}
\prod_{p=1}^m \left({\hat N}_c+p/m\right) \ . 
\quad {\hat N}_c={\hat c}^*{\hat c}  \ . \ \ \ \ \ 
\label{11}
\end{eqnarray}
The form (\ref{10}) for $m=2$ can be seen in Appendix of Ref.\citen{KPPYI}. 
The state $\ket{c_0}$ is an eigenstate of ${\hat \gamma}$ defined in the 
following :
\begin{equation}\label{12}
{\hat \gamma}\ket{c_0}=\gamma\ket{c_0} \ , \qquad
{\hat \gamma}=\sqrt{m^m}\sqrt{F_m^0({\hat N}_c)}\ {\hat c} \ .
\end{equation}
The form (\ref{10}) also shows that $\ket{c_0}$ is deformed from 
the conventional boson coherent state. However, the above treatment 
contains unsatisfactory point. The system is treated only in 
the subspace ${\wtilde B}_0$ of ${\wtilde B}$. 
Therefore, in order to complete the treatment, it may be necessary to 
investigate the effect coming from the subspace orthogonal to ${\wtilde B}_0$. 
The aim of this note is to investigate the above-mentioned problem.

For our present aim, we treat the system in the whole space ${\wtilde B}$. 
In this case, the orthogonal set consists of the following states : 
\begin{eqnarray}\label{13}
& &\kket{mn+r}=\left(\sqrt{(mn+r)!}\right)^{-1}
({\wtilde c}^*)^{mn+r}\kket{0} \ . \nonumber\\
& &\qquad\qquad\qquad
(n=0,1,2,\cdots ; \ r=0,1,2,\cdots, m-1)
\end{eqnarray}
Of course, the set $\{ \kket{mn+r} \}$ is complete and 
the space spanned by the states with $r=0$ 
corresponds to the subspace ${\wtilde B}_0$. 
The case $m=2$ was investigated in Ref.\citen{KPPYI} in a form different 
from the present one. 

In order to obtain the image of the space ${\wtilde B}$, let us 
prepare a boson space consisting of not only the boson 
$({\hat c} , {\hat c}^*)$ but also the boson $({\hat d} , {\hat d}^*)$, 
the orthogonal set of which is given by 
\begin{equation}\label{14}
\ket{n, r}=\ket{n}\otimes \rket{r} \ ,  \qquad\qquad\qquad\qquad\qquad\ 
\end{equation}
\vspace{-1cm}
\begin{subequations}\label{15all}
\begin{eqnarray}
& &\rket{r}=(\sqrt{r!})^{-1} ({\hat d}^*)^r \rket{0} \ , \qquad
(r=0, 1, 2, \cdots) 
\label{15}\\
& &{\hat d}\rket{0}=0 \ . 
\label{15a}
\end{eqnarray}
\end{subequations}
We call the above space as ${\overline B}$. The state $\ket{n}$ is defined in 
the relation (\ref{4}). We make the set $\{ \kket{mn+r} \}$ correspond 
to the set $\{ \ket{n, r} \}$ : 
\begin{equation}\label{16}
\kket{mn+r} \sim \ket{n, r}\ , \qquad \hbox{\rm i.e.,}\qquad
\ket{n,r}={\mib U}\kket{mn+r}\ .
\end{equation}
Since $r=0, 1, 2,\cdots, m-1$ in the set $\{ \kket{mn+r} \}$, $r$ in the 
state $\ket{n, r}$ shown in the relation (\ref{16}) is also 
equal to $0, 1, 2, \cdots, m-1$. 
This means that the set $\{ \ket{n, r} \}$ as the image of 
$\{ \kket{mn+r} \}$ composes a subspace of $\overline{B}$, which we 
denote ${\hat B}$. The operator ${\mib U}$ is written as 
\begin{equation}\label{17}
{\mib U}=\sum_{n=0}^\infty \sum_{r=0}^{m-1}
\ket{n, r}\bbra{mn+r} \ .
\end{equation}
The operator ${\mib U}$ satisfies the relation 
\begin{equation}\label{18}
{\mib U}^{\dagger}{\mib U}=1 \ , \qquad
{\mib U}{\mib U}^{\dagger}={\hat P}_m \ . \qquad
\left({\hat P}_m=\sum_{r=0}^{m-1}\rket{r}\rbra{r} \ \right)
\end{equation}

As a possible and natural extension of the multiboson coherent state 
$\kket{c_0}$, we define the following state in the space ${\wtilde B}$ : 
\begin{eqnarray}
& &\kket{c}=\left(\sqrt{\Gamma}\right)^{-1}
\sum_{n=0}^\infty \sum_{r=0}^{m-1}\left(\sqrt{(mn+r)!}\right)^{-1}
\gamma^n \delta^r \kket{mn+r} \ , 
\label{19}\\
& &\Gamma= \sum_{n=0}^\infty \sum_{r=0}^{m-1}
[(mn+r)!]^{-1} (|\gamma|^2)^n (|\delta|^2)^r \ .
\label{20}
\end{eqnarray}
Here, $\delta$ denotes a complex parameter additional to $\gamma$. 
If restricted to $r=0$ in the summation in (\ref{19}) and (\ref{20}), 
$\kket{c}$ is reduced to $\kket{c_0}$. 
With the use of ${\mib U}$ given in the relation (\ref{17}), the image 
of $\kket{c}$, which we denote $\ket{c}$, can be derived : 
\begin{eqnarray}
\ket{c}&=&{\mib U}\kket{c} \nonumber\\
&=&\left(\sqrt{\Gamma}\right)^{-1}\sum_{n=0}^{\infty}\sum_{r=0}^{m-1}
f_m(n, r)(\sqrt{n!})^{-1}(\sqrt{r!})^{-1}\gamma^n \delta^r \ket{n, r} \ , 
\label{21}\\
& &\hspace{-1cm}
f_m(n, r)=\sqrt{n!}\sqrt{r!}\left(\sqrt{(mn+r)!}\right)^{-1} \ .
\label{22}
\end{eqnarray}
The form (\ref{21}) shows that the state $\ket{c}$ is different from 
the conventional boson coherent state. The state $\ket{c}$ can be rewritten in 
the form 
\begin{eqnarray}
& &\ket{c}=\left(\sqrt{\Gamma}\right)^{-1}
\exp\left(\left(\sqrt{m^m}\right)^{-1} \gamma{\hat c}^*\left(
\sqrt{F_m({\hat N}_c, {\hat N}_d)}\right)^{-1}\right)
\cdot \exp\left(\delta{\hat d}^* {\hat P}_{m-1}\right)\ket{0} \ . 
\nonumber\\
& &\qquad\qquad\qquad\qquad\qquad\qquad\qquad\qquad\qquad\qquad\qquad
\qquad
(\ket{0}\otimes \rket{0}=\ket{0} )
\label{23}\\
& &F_m({\hat N}_c, {\hat N}_d)=\left({\hat N}_c+1\right)^{-1}
\prod_{p=1}^m \left[ {\hat N}_c+({\hat N}_d+p)/m\right] \ , 
\qquad {\hat N}_d={\hat d}^*{\hat d} \ , 
\label{24}\\
& &{\hat P}_{m-1}=\sum_{r=0}^{m-2} \rket{r}\rbra{r} \ . 
\label{25}
\end{eqnarray}
The function $F_m({\hat N}_c, {\hat N}_d)$ is a natural extension from 
$F_m^0({\hat N}_c)$ shown in the relation (\ref{11}). 
The part $\exp(\delta{\hat d}^*{\hat P}_{m-1})\ket{0}$ is a 
concrete example of the case investigated in Ref.\citen{KPPTYIII}.

The state $\ket{c}$ has a close connection with the following operators : 
\begin{eqnarray}
& &{\hat \gamma}=\sqrt{m^m}\sqrt{F_m({\hat N}_c, {\hat N}_d)}\ {\hat c} \ , 
\label{26}\\
& &{\hat \delta}=\sqrt{(m{\hat N}_c+{\hat N}_d+1)\cdot({\hat N}_d+1)^{-1}}
\ {\hat d} \ . 
\label{27}
\end{eqnarray}
The connection can be seen in the relation 
\begin{eqnarray}
& &{\hat \gamma}\ket{c}=\gamma\ket{c} \ , 
\label{28}\\
& &{\hat \delta}\ket{c}=\delta{\hat P}_{m-1}\ket{c} \ . 
\label{29}
\end{eqnarray}
%For the proof of the above relation, the following relation is useful : 
It is useful to note that the following relation for $F_m$ is satisfied : 
\begin{equation}\label{30}
F_m({\hat N}_c, {\hat N}_d+1)
=\left[{\hat N}_c+1+({\hat N}_d+1)/m\right]\left[
{\hat N}_c+({\hat N}_d+1)/m\right]^{-1}F_m({\hat N}_c, {\hat N}_d) \ .\ \ 
\end{equation}
With use of the relation (\ref{30}), we can prove 
\begin{equation}\label{31}
\left[ {\hat \delta} , {\hat c}^*\left(
\sqrt{F_m({\hat N}_c, {\hat N}_d)}\right)^{-1} \right]
=0 \ .
\end{equation}
Further, we have 
\begin{equation}\label{32}
\left[ {\hat \gamma} , {\hat c}^*\left(
\sqrt{F_m({\hat N}_c, {\hat N}_d)}\right)^{-1} \right]
=1 \ .
\end{equation}
The interpretation of the relation (\ref{29}) is given in Ref.\citen{KPPTYIII}. 

Our present system is also considered in terms of the boson type classical 
canonical variables. 
The canonicity condition in the TDHF theory\cite{MMSK} enables us 
to perform this task. We introduce two sets of the variables 
$(c, c^*)$ and $(d, d^*)$. In the same manner as that in (I), 
we can show that $c$ and $d$  are related with 
\begin{equation}\label{33}
c=\gamma\sqrt{\frac{\partial \Gamma}{\partial |\gamma|^2}\cdot \Gamma^{-1}} 
\ , 
\qquad
d=\delta\sqrt{\frac{\partial \Gamma}{\partial |\delta|^2}\cdot \Gamma^{-1}} \ .
\end{equation}
The expectation values of ${\hat N}_c$ and ${\hat N}_d$ for $\ket{c}$, 
which we denote $N_c$ and $N_d$, respectively, are given as 
\begin{subequations}\label{34}
\begin{eqnarray}
N_c&=&
\bra{c}{\hat \gamma}^*\left(\sqrt{F_m({\hat N}_c, {\hat N}_d)}\right)^{-1}
(\sqrt{m^m})^{-1}\cdot (\sqrt{m^m})^{-1}
\left(\sqrt{F_m({\hat N}_c, {\hat N}_d)}
\right)^{-1}{\hat \gamma}\ket{c} \nonumber\\
&=&|\gamma|^2\bra{c} (m^m)^{-1} F_m({\hat N}_c, {\hat N}_d)^{-1}\ket{c} 
\nonumber\\
&=&|\gamma|^2\cdot \frac{\partial \Gamma}{\partial |\gamma|^2}\Gamma^{-1} \ , 
\label{34a}\\
N_d&=&
\bra{c}{\hat \delta}^*\sqrt{(m{\hat N}_c+{\hat N}_d+1)^{-1}
({\hat N}_d+1)}\cdot
\sqrt{(m{\hat N}_c+{\hat N}_d+1)^{-1}
({\hat N}_d+1)}\ {\hat \delta}\ket{c} \nonumber\\
&=&|\delta|^2\bra{c} {\hat P}_{m-1}({\hat N}_d+1)(m{\hat N}_c+{\hat N}_d
+1)^{-1} {\hat P}_{m-1}\ket{c} 
\nonumber\\
&=&|\delta|^2\cdot \frac{\partial \Gamma}{\partial |\delta|^2}\Gamma^{-1} \ . 
\label{34b}
\end{eqnarray}
\end{subequations}
The last line of each equation is directly derived from 
$\bra{c}{\hat c}^*{\hat c}\ket{c}$ using (\ref{21}). 
Thus, we have
\begin{equation}\label{35}
N_c=c^* c \ , \qquad N_d=d^* d\ , \qquad\qquad\qquad\qquad\qquad\qquad\quad
\end{equation}
\vspace{-0.9cm}
\begin{subequations}\label{36}
\begin{eqnarray}
& &\frac{\partial \Gamma}{\partial |\gamma|^2}\Gamma^{-1}
=\bra{c} (m^m)^{-1} F_m({\hat N}_c , {\hat N}_d)^{-1} \ket{c} \ , 
\label{36a}\\
& &\frac{\partial \Gamma}{\partial |\delta|^2}\Gamma^{-1}
=\bra{c} {\hat P}_{m-1}({\hat N}_d+1)(m{\hat N}_c+{\hat N}_d
+1)^{-1} {\hat P}_{m-1}\ket{c} \ . 
\label{36b}
\end{eqnarray}
\end{subequations}
The relations (\ref{33}) and (\ref{36}) lead us to 
\begin{subequations}\label{37}
\begin{eqnarray}
& &\gamma=\left[\sqrt{\bra{c} (m^m)^{-1} F_m({\hat N}_c , {\hat N}_d)^{-1} 
\ket{c}}\right]^{-1} c \ , 
\label{37a}\\
& &\delta=\left[\sqrt{\bra{c} {\hat P}_{m-1}({\hat N}_d+1)
(m{\hat N}_c+{\hat N}_d+1)^{-1} {\hat P}_{m-1}\ket{c}}\right]^{-1} d \ . 
\label{37b} 
\end{eqnarray}
\end{subequations}
For the relations (\ref{37}), we make a rough approximation : 
\begin{eqnarray}\label{38}
& &{\hat P}_{m-1}\ket{c} \approx \ket{c} \ , \nonumber\\
& &\bra{c} G({\hat N}_c, {\hat N}_d) \ket{c} \approx G(N_c, N_d) \ .
\end{eqnarray}
Then, we have approximate forms for $\gamma$ and $\delta$ as follows : 
\begin{subequations}\label{39}
\begin{eqnarray}
& &\gamma=\sqrt{m^m}\sqrt{F_m(N_c, N_d)}\ c \ , 
\label{39a}\\
& &\delta=\sqrt{(mN_c+N_d+1)(N_d+1)^{-1}}\ d \ .
\label{39b}
\end{eqnarray}
\end{subequations}
With the use of the relation (\ref{37}), we can give the 
expectation values of various operators for $\ket{c}$ in terms of 
$(c, c^*)$ and $(d, d^*)$ in the approximate forms. 

The above is an outline of our idea for completing the treatment 
of the mutiboson coherent state given in (I). 
Finally, we give some remarks. In this paper, the boson operator 
$({\hat d}, {\hat d}^*)$ was used for the completion. 
Instead of the boson operator, we can use the operator 
$({\maru d}, {\maru d}^*)$ which obeys the commutation relation 
\begin{equation}\label{40}
\left[ {\maru d} , {\maru d}^* \right]
=1 - \frac{m}{(m-1)!} \left({\maru d}^*\right)^{m-1}
\left({\maru d}\right)^{m-1} \ . 
\end{equation}
Under the presupposition of the existence of the state $\rket{0}$ 
obeying ${\maru d}\rket{0}=0$, the relation (\ref{40}) leads us 
automatically to $({\maru d}^*)^m \rket{0}=0$. 
This means that we can formulate our idea without worrying the 
existence of the subspace ${\hat B}$. 
Actually, the case $m=2$ was treated by the present authors.\cite{KPPYII} 
In this case, the relation (\ref{40}) gives us 
${\maru d}{\maru d}^*+{\maru d}^*{\maru d}=1$ and 
$({\maru d}, {\maru d}^*)$ can be regarded as fermion operator. 
In Ref.\citen{KPPYII}, we adopted the coherent state different from 
the present one and, then, it is impossible to compare the results 
directly. The detail of the relation (\ref{40}) was discussed by 
the present authors\cite{KPTY} in relation to the 
Dirac canonical theory for constraint systems.\cite{D}

%\section*{Acknowledgements}


\begin{thebibliography}{99}
%%%%%%%%%%%%%%%%%%%%%%%%%%%%%%%%%%%%%%%%%%%%%%%%%%%%%%%%%%%%%
% Some macros are available for the bibliography:
%   o for general use
%      \JL : general journals          \andvol : Vol (Year) Page
%   o for individual journal 
%      \PR  : Phys. Rev.               \PRL : Phys. Rev. Lett.
%      \NP  : Nucl. Phys.              \PL  : Phys. Lett.
%      \JMP : J. Math. Phys.           \CMP : Commun. Math. Phys.
%      \PTP : Prog. Theor. Phys.       \JPSJ: J. Phys. Soc. Jpn.
%      \JP  : J. of Phys.              \NC  : Nouvo Cim.
%      \IJMP: Int. J. Mod. Phys.       \ANN : Ann. of Phys.
% Usage:
%   \PR{D45,1990,345}            ==> Phys.~Rev.\ {\bf D45} (1990), 345
%   \JL{Phys.~Lett.,A30,1981,56} ==> Phys.~Lett.\ {\bf A30} (1981), 56
%   \andvol{B123,1995,1020}      ==> {\bf B123} (1995), 1020
%%%%%%%%%%%%%%%%%%%%%%%%%%%%%%%%%%%%%%%%%%%%%%%%%%%%%%%%%%%%%
\bibitem{KPPTYI}
A. Kuriyama, C. Provid\^encia, J. da Provid\^encia, Y. Tsue and M.~Yamamura, 
        Prog.~Theor.~Phys.~{\bf 106} (2001), 751.  
\bibitem{KPPTYII}
A. Kuriyama, C. Provid\^encia, J. da Provid\^encia, Y. Tsue and M.~Yamamura, 
        Prog.~Theor.~Phys.~{\bf 106} (2001), 765. \\
A. Kuriyama, C. Provid\^encia, J. da Provid\^encia, Y. Tsue and M.~Yamamura, 
        To appear in Prog.~Theor.~Phys. {\bf 107} (2002), No. 1. 
%        (nucl-th/0110032).
\bibitem{KPPTYIII}
A. Kuriyama, C. Provid\^encia, J. da Provid\^encia, Y. Tsue and M.~Yamamura, 
        To appear in Prog.~Theor.~Phys. {\bf 107} (2002), No. 2.
%         (nucl-th/0111029 ).
\bibitem{Penson}
K. A. Penson and A. I. Solomon, J. Math. Phys. {\bf 40} (1999), 
2354.
\bibitem{KPPYI}
A. Kuriyama, C. Provid\^encia, J. da Provid\^encia and M.~Yamamura, 
        Prog.~Theor.~Phys.~{\bf 103} (2000), 733.  
\bibitem{KPPYII}
A. Kuriyama, C. Provid\^encia, J. da Provid\^encia and M.~Yamamura, 
        Prog.~Theor.~Phys.~{\bf 104} (2000), 155.  
\bibitem{MYT}
T. Marumori, M. Yamamura and A. Tokunaga, Prog. Theor. Phys. {\bf 31} 
(1964), 1009. 
\bibitem{MMSK}
T. Marumori, T. Maskawa, F. Sakata and A. Kuriyama, 
Prog. Theor. Phys. {\bf 64} (1980), 1294. \\
M. Yamamura and A. Kuriyama, Prog. Theor. Phys. Suppl. No. 93 (1987), 1. 
\bibitem{KPTY}
A. Kuriyama, J. da Provid\^encia, Y. Tsue and M.~Yamamura, 
        Prog.~Theor.~Phys.\ {\bf 95} (1996), 79. 
\bibitem{D}
P. A. M. Dirac, Can. J. Math. {\bf 21} (1950), 129.
\end{thebibliography}
\end{document}